\documentclass[aip,prb,twocolumn,showpacs]{revtex4}
\usepackage{graphicx,color}
\usepackage{amsmath}
\usepackage{longtable}
\begin{document}

\title{Practical thermodynamics of Yukawa systems at strong coupling}

\author{Sergey A. Khrapak,$^{1,2,}$\footnote{Also at Joint Institute for High Temperatures, Russian Academy of Sciences, Moscow, Russia} Nikita P. Kryuchkov,$^3$ Stanislav O. Yurchenko,$^3$ and Hubertus M. Thomas$^1$}
\date{\today}
\affiliation{$^1$Forschungsgruppe Komplexe Plasmen, Deutsches Zentrum f\"{u}r Luft- und Raumfahrt,
Oberpfaffenhofen, Germany \\$^2$Aix-Marseille-Universit\'{e}, CNRS, Laboratoire PIIM, UMR 7345, 13397
Marseille cedex 20, France \\$^3$Bauman Moscow State Technical University, 2-nd Baumanskaya str. 5, Moscow 105005, Russia}

\begin{abstract}
Simple practical approach to estimate thermodynamic properties of strongly coupled Yukawa systems, in both fluid and solid phases, is presented. The accuracy of the approach is tested by extensive comparison with direct computer simulation results (for fluids and solids) and the recently proposed shortest-graph method (for solids). Possible applications to other systems of softly repulsive particles are briefly discussed.
\end{abstract}

\maketitle

\section{Introduction}

A large class of soft matter systems can be described generically as massive charged particles immersed in a neutralizing medium of lighter particles.~\cite{Hopkins}
Two well-known examples are colloidal suspensions~\cite{Lowen,Morrison,IvlevBook} and complex (dusty) plasmas.~\cite{IvlevBook,FortovRev,ShuklaRMP,FortovBook,ChaudhuriSM}
The interaction between the massive particles is a key issue to adequately understand most of the phenomena in these systems. Although it can be rather complex in real situations, it is often assumed that the electrical interactions between the particles can be described, in the zero approximation, by the Yukawa (also known as the screened Coulomb or Debye-H\"{u}ckel) repulsive potential
\begin{equation}\label{Yukawa}
V(r)= (Q^2/r)\exp(-r/\lambda_{\rm D}),
\end{equation}
where $Q$ is the particle charge, $\lambda_{\rm D}$ is the Debye screening length associated with the neutralizing medium, and $r$ is the distance between a pair of particles.
The effect of the neutralizing medium on the effective interactions between the particles can involve more than only screening. This is particularly relevant in complex plasmas, where continuous absorption of plasma electrons and ions on the particle surface result in inverse-power-law asymptotes of interaction at large interparticle separations.~\cite{FortovRev,ChaudhuriSM,KhrapakPRL2008,KhrapakCPP,Semenov} Plasma production and loss processes can also produce a double-Yukawa interaction potential characterized by two different screening lengths.~\cite{KhrapakPoP2010,Wysocki} Nevertheless, many experimentally observed trends can be already reproduced by the simplest model considering point-like particles interacting via the repulsive Yukawa potential (\ref{Yukawa}), at least qualitatively. For this reason this model, which we will refer to as the single component Yukawa system, has received a great deal of interest. Another remarkable property, which makes the model Yukawa interaction particularly important for the soft condensed matter research, is that by varying $\lambda_{\rm D}$ an extremely broad range of interaction steepness can be explored. The two limiting cases correspond to extremely soft unscreened Coulomb interaction (one-component-plasma limit) for $\lambda_{\rm D}\rightarrow \infty$ and extremely steep hard-sphere-like repulsion for $\lambda_{\rm D}\rightarrow 0$.

Depending on the strength of the interparticle interaction, Yukawa systems can be regarded as weakly or strongly coupled. Weak coupling corresponds to the situation when the energy of the interaction at a characteristic mean interparticle separation $\Delta$ is much smaller than the system temperature $T$ (expressed in energy units), so that $V(\Delta)\ll T$. In this regime the interaction provides only small corrections to the thermodynamic quantities of an ideal gas. These corrections can be relatively easy evaluated (e.g. by means of the second virial coefficient). In the opposite regime, $V(\Delta)\gg 1$, the interaction produce significant (or dominant) contribution to the thermodynamic quantities. The particles form either the fluid phase or, at even stronger coupling, the solid phase. Evaluation of thermodynamic properties becomes less trivial than in the limit of weak coupling. The focus of the present paper is, therefore, on the strong coupling regime.

Thermodynamic properties and phase diagram of strongly coupled Yukawa systems are relatively well investigated using various computational and analytical techniques. Some relevant examples include Monte Carlo (MC) and molecular dynamics (MD) numerical simulations,~\cite{Robbins,Meijer,Hamaguchi96,Hamaguchi97,CG} as well as integral equation theoretical studies~\cite{Tejero,Kalman2000,Faussurier,Gapinski,Hlushak,SMSA} (see also references therein).~\cite{Com1} In particular, accurate results for the excess internal energy obtained in MD and MC simulations have been tabulated in Refs.~\onlinecite{Hamaguchi97,CG} for a number of state points in the phase diagram of Yukawa systems. However, since the calculation of other thermodynamic functions requires knowledge of certain integrals and derivatives of the excess energy (see examples below), these data are very useful as reference points, but less appropriate as a practical tool for estimating thermodynamic properties. Semi-empirical fitting formulas~\cite{TotsujiJPA,TotsujiPoP,Vaulina2010} and simplistic approaches~\cite{DHH,ISM} to estimate thermodynamics of Yukawa systems have been also discussed in the literature. These can be useful in certain situations, but are not very helpful when sufficient accuracy is required.

The purpose of the present paper is to describe simple practical approach to evaluate thermodynamic properties of Yukawa systems at strong coupling, {\it both in the fluid and solid phases}. The approach is based on simple phenomenological arguments, which can be also relevant to a wider class of soft repulsive interactions. The accuracy of the approach is tested by calculating the compressibility factor and comparing this with available as well as new results from direct computer simulations. We also use the recently proposed shortest-graph method,~\cite{Yurchenko,YurchenkoPRB} which allows to calculate the radial distribution function and, hence, pressure in crystals. The relative deviations between theoretical and simulation results normally do not exceed one part in one thousand for dense fluids not too far from crystallization and solids not too close to the melting point. The largest deviations are documented for the solid phase near melting transition and superheated solid, but even there they are below $1\%$. Hence, the proposed approach represents very convenient and accurate tool to estimate thermodynamics of Yukawa and other related systems in a very broad parameter regime.

\section{Thermodynamic functions of Yukawa systems}\label{TF}

The main thermodynamic quantities of interest here are the internal energy $U$, Helmholtz free energy $F$, and pressure $P$ of the system of $N$ Yukawa particles, occupying volume $V$ and having temperature $T$. These thermodynamic functions are related via~\cite{Landau}
\begin{eqnarray}
U=-T^2\left(\frac{\partial}{\partial T}\frac{F}{T}\right)_V, \\
P=-\left(\frac{\partial F}{\partial V}\right)_T.
\end{eqnarray}
We will use conventional reduced units: $u=U/NT$, $f=F/NT$, and $p=PV/NT$. The latter ratio, $Z=PV/NT$, is also known as the compressibility factor.

The single component Yukawa system is conventionally characterized by two dimensionless parameters. The first is the coupling parameter, $\Gamma=Q^2/aT$, where $a= (3V/4\pi N)^{1/3}$ is the Wigner-Seitz radius and the temperature $T$ is expressed in energy units throughout this paper. The second is the screening parameter, $\kappa=a/\lambda_{\rm D}$.
It is useful to express reduced thermodynamic functions in terms of Yukawa system phase state variables, $\kappa$ and $\Gamma$. For a fixed number of particles we have $\Gamma\propto (aT)^{-1}\propto V^{-1/3}T^{-1}$ and $\kappa\propto a\propto V^{1/3}$. This implies
\begin{displaymath}
\frac{\partial \Gamma}{\partial T}=-\frac{\Gamma}{T}, \quad \frac{\partial \Gamma}{\partial V}=-\frac{1}{3}\frac{\Gamma}{V}, \quad \frac{\partial \kappa}{\partial T}= 0, \quad \frac{\partial \kappa}{\partial V}=\frac{1}{3}\frac{\kappa}{V}.
\end{displaymath}
The following equations for the reduced thermodynamic functions are then obtained:
\begin{equation}\label{f_def}
\frac{u}{\Gamma}=\frac{\partial f}{\partial \Gamma},
\end{equation}
and
\begin{equation}\label{Z_def}
Z=1+\frac{\Gamma}{3}\frac{\partial f}{\partial \Gamma}-\frac{\kappa}{3}\frac{\partial f}{\partial \kappa}.
\end{equation}

To conclude this section we remind that in this paper we consider exclusively the single component Yukawa system, where thermodynamics is completely determined by particle-particle interactions and correlations. In real systems (e.g. complex plasmas or colloidal suspensions) neutralizing medium is normally present to neutralize and stabilize like-charged particles. In this case, particle-medium interactions also affect thermodynamics. However, the effect of particle-medium interactions is additive to that of particle-particles interactions and can be easily evaluated.~\cite{Ham94} For example, the excess (free) energy associated with the presence of neutralizing medium (e.g. plasma) is
\begin{equation}
f_{\rm pl}=u_{\rm pl}=-\frac{3\Gamma}{2\kappa^2}-\frac{\kappa\Gamma}{2}.
\end{equation}
Note that the contribution of the neutralizing medium is negative and dominant at strong coupling, implying that the excess energy and pressure of the corresponding system are also negative in this regime. For the single component Yukawa system these quantities are obviously positive.

\section{Yukawa solids}

The reduced excess energy of a solid in the harmonic approximation is
\begin{equation}
u_{\rm s}=M_{\rm s}\Gamma+\tfrac{3}{2},
\end{equation}
where $M_{\rm s}$ is the corresponding lattice sum (Madelung constant). The reduced Helmholtz free energy of a solid is related to the excess energy by the standard integration [Eq.~(\ref{f_def})], which yields
\begin{equation}
f_{\rm s}=M_{\rm s}(\Gamma-\Gamma_{\rm m})+\tfrac{3}{2}\ln(\Gamma/\Gamma_{\rm m})+f_{\rm m},
\end{equation}
where the integration starts from $\Gamma_{\rm m}$ -- the coupling parameter at the fluid-solid phase transition, and $f_{\rm m}=f(\Gamma_{\rm m})$. According to the Ross melting criterion,~\cite{Ross} which he obtained by reformulating the celebrated Lindemann melting law in terms of the statistical-mechanical partition function
\begin{equation}\label{f_m}
f_{\rm m}\simeq M_{\rm s}\Gamma_{\rm m} +{\mathcal C},
\end{equation}
where ${\mathcal C}$ is some constant. In other words, the thermal component of the reduced excess free energy remains {\it approximately} constant at melting. Hoover {\it et al}.~\cite{Hoover} observed that ${\mathcal C}\simeq 6$ for several inverse-power-law (IPL) repulsive potentials. More recently, Agrawal and Kofke~\cite{Agrawal} documented some dependence of ${\mathcal C}$ on the potential softness (${\mathcal C}$ somewhat increases for softer potentials) for a wide range of IPL potentials. However, this variation remains relatively weak, with $5.9\lesssim {\mathcal C}\lesssim 6.6$ in a very wide range of softness investigated.

\begin{figure}
\includegraphics[width=7.5cm]{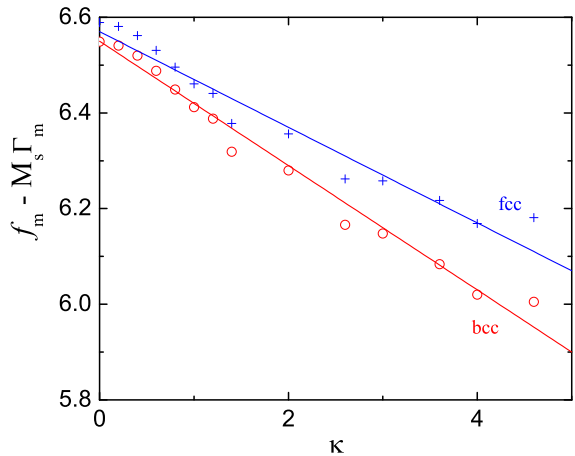}
\caption{(Color online) Thermal component of the reduced excess free energy of Yukawa solids at melting. Red circles correspond to the solid bcc lattice, blue crosses denote the solid fcc phase. Symbols are calculated from the harmonic lattice approximation using the data tabulated in Refs.~\onlinecite{Hamaguchi96,Hamaguchi97}. Solid lines are linear fits to these data points.}
\label{fig1}
\end{figure}

We have calculated the values of ${\mathcal C}$ at melting of the bcc and fcc Yukawa solids, in the harmonic lattice approximation, using the data tabulated in Refs.~\onlinecite{Hamaguchi96,Hamaguchi97}. In addition to neglecting anharmonic corrections, we assumed that fluid-bcc and fluid-fcc transition lines are very close to each other in the parameter regime of present interest (for the confirmation see Fig.~7 of Ref.~\onlinecite{Tejero}) so that a unique value of $\Gamma_{\rm m}$ can be used (this is further justified by the fact that $\Gamma_{\rm m}$ appears under the logarithm when calculating ${\mathcal C}$). The results are shown in Fig.~\ref{fig1}. Taking into account some scattering in the data points, it is reasonable to apply linear fitting procedure. This yields
\begin{equation}\label{C}
{\mathcal C}(\kappa) = \begin{cases} 6.55-0.13\kappa,  & {\rm (bcc~~~ lattice)} \\ 6.57-0.10\kappa & {\rm (fcc~~~~ lattice)}.  \end{cases}
\end{equation}

The resulting expression for the reduced free energy of the solid phase is
\begin{equation}\label{f_sol}
f_{\rm s}(\kappa,\Gamma)=M_{\rm s}(\kappa)\Gamma+\tfrac{3}{2}\ln\left[\Gamma/\Gamma_{\rm m}(\kappa)\right]+{\mathcal C}(\kappa).
\end{equation}
To use Eq.~(\ref{f_sol}) for practical evaluations we need to specify the dependence $\Gamma_{\rm m}(\kappa)$. A simple expression proposed by Vaulina {\it et al}.~\cite{VaulinaJETP,VaulinaPRE} is suitable for this purpose~\cite{PractUP}
\begin{equation}\label{melt}
\Gamma_{\rm m}(\kappa)\simeq \frac{172 \exp(\alpha\kappa)}{1+\alpha\kappa+\tfrac{1}{2}\alpha^2\kappa^2},
\end{equation}
where the constant $\alpha=(4\pi/3)^{1/3}\simeq 1.612$  is the ratio of the mean interparticle distance $\Delta=(V/N)^{1/3}$ to the Wigner-Seitz radius $a$.

\begin{table}
\caption{\label{Tab1} The compressibility factor (reduced pressure)  for the single component Yukawa fcc solid. The first two columns specify the location of the system state point on the $(\kappa,\Gamma)$ plane. The third column lists the values of the reduced coupling strength $\Gamma/\Gamma_{\rm m}$, as calculated using Eq.~(\ref{melt}). (Note that the last three rows may correspond to a superheated solid). The remaining columns contain the values of $Z$ obtained using MC simulations by Meijer and Frenkel~\cite{Meijer} and tabulated in Ref.~\onlinecite{Tejero} ($Z_{\rm MC}$), present theoretical approach ($Z_{\rm present}$), and the shortest-graph method~\cite{Yurchenko,YurchenkoPRB} ($Z_{\rm SG}$). }
\begin{ruledtabular}
\begin{tabular}{lllllll}
$\kappa$ & $\Gamma$ & $\Gamma/\Gamma_{\rm m}$ &  $Z_{\rm MC}$ & $Z_{\rm present}$ & $Z_{\rm SG}$ \\ \hline
4.034 & 5764.8 & 1.44 & 41.058 & 40.941 & 41.197  \\
4.066 & 5719.1 & 1.37 & 38.927 & 38.809 & 39.075  \\
4.096 & 5678.3 & 1.32 & 37.115 & 36.992 & 37.274  \\
4.128 & 5634.0 & 1.26 & 35.259 & 35.096 & 35.399  \\
4.169 & 5577.8 & 1.19 & 32.969 & 32.820 & 33.126  \\
4.208 & 5526.8 & 1.12 & 31.005 & 30.870 & 31.202  \\
4.251 & 5470.4 & 1.05 & 28.991 & 28.883 & 29.196  \\
4.300 & 5408.0 & 0.98 & 26.919 & 26.729 & 27.113  \\
4.344 & 5353.2 & 0.92 & 25.211 & 25.000 & 25.409  \\
4.406 & 5277.8 & 0.84 & 22.996 & 22.799 & 23.268  \\
\end{tabular}
\end{ruledtabular}
\end{table}

Equation~(\ref{Z_def}) complemented by Eqs.~(\ref{C})--(\ref{melt}) allows us to calculate the compressibility of Yukawa solids. We compare our theoretical results with the direct MC simulation of Meijer and Frenkel.~\cite{Meijer} In addition, the compressibility has been evaluated using the recently proposed {\it shortest-graph} method.~\cite{Yurchenko,YurchenkoPRB} In this method the pair distribution function $g({\bf r})$ is calculated by assuming that it can be represented as a sum of the Gaussian peaks which describe the corresponding partial contributions of the particles in the crystal. The radial distribution function is then obtained by angular integration, $g(r)=(1/4\pi)\int d\Omega g({\bf r})$. The shortest-graph method allows to calculate $g(r)$  for a wide range of interaction potentials, crystalline lattices, temperatures, and densities.~\cite{Yurchenko,YurchenkoPRB} The compressibility is obtained from
the virial equation of state,~\cite{HansenBook}
\begin{displaymath}
Z=1-\frac{2\pi N}{3VT}\int_0^{\infty}r^3V'(r)g(r)dr.
\end{displaymath}
The details of implementation of the shortest-graph method are described in Ref.~\onlinecite{YurchenkoPRB}.

Numerical and theoretical results for the compressibility of the fcc solid phase are summarized in Table~\ref{Tab1}. Our present theoretical results are always below the MC results, the relative difference increases on approaching melting and, in particular, in the superheated regime. However, even in this case the difference is below $1\%$.
The shortest-graph method demonstrates comparable accuracy, but overestimates the compressibility obtained in MC simulations. The documented accuracy is considerably better than (for example) the hard-sphere perturbation theory described in Ref.~\onlinecite{Tejero} can provide.

Since the data tabulated previously~\cite{Tejero} are limited to a relatively narrow range of $\kappa$ around $\kappa\simeq 4$ and are all for the fcc solid phase, we performed additional comparison   for the regime of smaller $\kappa$, where Yukawa solid forms a stable bcc lattice. The MD simulations have been performed for $N=64 000$ particles in the canonical ($NVT$) ensemble, using the Langevin thermostat. The numerical time step is set to $10^{-4} \sqrt{m\lambda_D^3/Q^2}$, where $m$ is the particle mass. The cutoff radius for the potential is $15\Delta$.
To obtain the equilibrium state and calculate the compressibility, the simulations have been run for $1.5\times 10^6$ steps.

\begin{table}
\caption{\label{Tab1_2} The compressibility factor (reduced pressure)  for the single component Yukawa bcc solid. The first three columns are similar to those in Tab~\ref{Tab1}. The remaining columns contain the values of $Z$ obtained using our MD simulation ($Z_{\rm MD}$), the shortest-graph method~\cite{Yurchenko,YurchenkoPRB} ($Z_{\rm SG}$) and the present theoretical approach ($Z_{\rm present}$). }
\begin{ruledtabular}
\begin{tabular}{lrlrrr}
$\kappa$ & $\Gamma$ & $\Gamma/\Gamma_{\rm m}$ &  $Z_{\rm MD}$ & $Z_{\rm SG}$ & $Z_{\rm present}$ \\ \hline
0.6 & 280.6 & 1.51 & 1090.711 & 1090.889 & 1090.340  \\
1.0 & 326.1 & 1.48 & 404.169 & 404.297  &  403.916  \\
1.4 & 403.2 & 1.42 & 213.525 & 213.637 & 213.410   \\
2.0 & 660.1 & 1.44 & 119.450 & 119.447 & 119.401  \\
2.7 & 1266.4 & 1.41 & 73.253 & 73.263 & 73.290   \\
4.0 & 5755.5 & 1.50 & 43.258 & 43.272 & 43.245   \\
\end{tabular}
\end{ruledtabular}
\end{table}

Comparison between MD simulations, the shortest-graph method, and the present theory is summarized in Table~\ref{Tab1_2}. The calculations using the shortest-graph method yield values extremely close to the ``exact'' MD results. The present analytical approach has almost the same accuracy, the agreement with MD results is very impressive. The relative deviations are well below one part in one thousand for all phase states. Reasons for these deviations can include the approximate character of the dependence $\Gamma_{\rm m}(\kappa)$ and the neglect of anharmonic corrections to the free energy. These are not easy to estimate. Given the simplicity of the present theoretical approach, the agreement should be considered as excellent. Thus, it can serve as a simple practical tool to evaluate thermodynamic functions of Yukawa solids when extreme accuracy is not required.

\section{Yukawa fluids near freezing}

The reduced excess energy of the fluid phase can be conveniently divided into static and thermal components~\cite{Rosenfeld1998,Rosenfeld2000,PractUP}
\begin{equation}\label{div}
u_{\rm fl}=M_{\rm fl}\Gamma+u_{\rm th}
\end{equation}
The first term corresponds to the static contribution and the coefficient $M_{\rm fl}$ is referred to as the fluid Madelung constant.~\cite{Rosenfeld2000}
It can be obtained with the Percus-Yevick (PY) radial distribution function of hard spheres in the limit $\eta =1$, where $\eta$ is the hard sphere packing fraction.~\cite{Rosenfeld1998,Rosenfeld2000} Alternatively it can be evaluated using the ion sphere model (ISM), where each particle is placed in the center of the charge neutral Wigner-Seitz spherical cell and the energy is then calculated from simple electrostatic consideration.~\cite{ISM} The result is
\begin{equation}
M_{\rm fl}(\kappa) = \frac{\kappa(\kappa+1)}{(\kappa+1)+(\kappa-1)e^{2\kappa}},
\end{equation}
The thermal component of the internal energy exhibits quasi-universal scaling for a wide class of soft repulsive potentials, as first pointed out by Rosenfeld and Tarazona (RT scaling).~\cite{Rosenfeld1998,Rosenfeld2000} The accuracy of this scaling for various model systems has been recently investigated in extensive numerical simulations.~\cite{Ingebrigtsen}
Previously, a variant of the RT scaling has been successively used to obtain practical expressions for the internal energy and pressure of Yukawa fluids across coupling regimes.~\cite{PractUP}
Here we modify this approach with the main emphasis on the near-freezing regime.

\begin{figure}
\includegraphics[width=7.5cm]{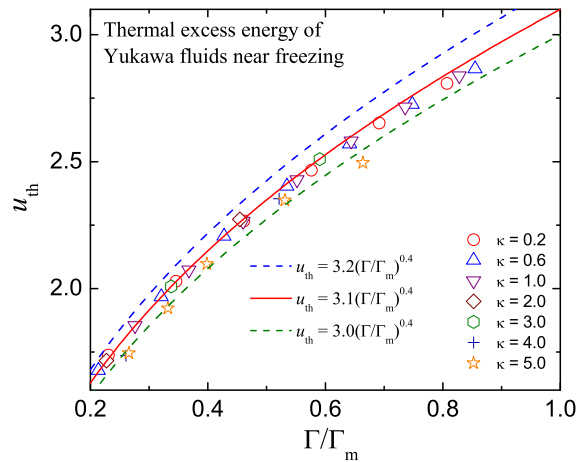}
\caption{(Color online) Thermal component of the reduced excess energy of Yukawa fluids near freezing versus the reduced coupling parameter $\Gamma/\Gamma_{\rm m}$. Symbols correspond to the numerical simulations for different values of the screening parameter $\kappa$.~\cite{Hamaguchi97,Farouki} The  curves correspond to the functional form $u_{\rm th}=\delta(\Gamma/\Gamma_{\rm m})^{2/5}$ with different values of $\delta$. The value $\delta\simeq 3.1$ is the most appropriate in the considered near-freezing regime.}
\label{fig2}
\end{figure}

For strongly coupled Yukawa fluids near freezing the RT scaling is illustrated in Fig.~\ref{fig2}, where the dependence of $u_{\rm th}$ on $\Gamma/\Gamma_{\rm m}$ is plotted for a number of screening parameters $\kappa<5$. Except the strong screening regime with $\kappa=5$ all other data points have a tendency to group around a single quasi-universal curve. Reasonably accurate fits (shown by the curves) can be obtained using the functional form
\begin{equation}\label{fit}
u_{\rm th}=\delta (\Gamma/\Gamma_{\rm m})^{2/5}.
\end{equation}
Originally, Rosenfeld suggested~\cite{Rosenfeld2000} to use $\delta=3.0$, while in our previous paper~\cite{PractUP} we used a modified expression, $u_{\rm th}=3.2 (\Gamma/\Gamma_{\rm m})^{2/5}-0.1$, in a wider range of coupling. Figure~\ref{fig2} demonstrates that a simple form (\ref{fit}) with $\delta=3.1$ is more appropriate near freezing and we use this value in calculations.

Integrating the excess energy from $\Gamma$ to $\Gamma_{\rm m}$ and subtracting this from the known value of $f_{\rm m}(\kappa)$ we get the free energy of Yukawa fluids in the form
\begin{eqnarray}\label{f_fl}
f_{\rm fl}(\kappa,\Gamma)= M_{\rm fl}(\kappa)\Gamma+\Gamma_{\rm m}(\kappa)\left[M_{\rm s}(\kappa)-M_{\rm fl}(\kappa)\right]+ \nonumber \\
\frac{5\delta}{2}\left[\left(\frac{\Gamma}{\Gamma_{\rm m}(\kappa)}\right)^{2/5}-1\right]+{\mathcal C}(\kappa).
\end{eqnarray}
Note that the RT scaling in the form of Eq.~(\ref{fit}) with constant $\delta$ only holds for $\kappa\lesssim 4$ while the fcc-bcc-fluid triple point is located near $\kappa\simeq 4.5$~\cite{Hamaguchi97,Hoy,KhrapakEPL2012} (bcc lattice being thermodynamically favorable at weak screening). Therefore, $M_{\rm bcc}(\kappa)$ and ${\mathcal C}_{\rm bcc}(\kappa)$ should be substituted into Eq.~(\ref{f_fl}) in the domain of its applicability.

We calculated the compressibility factors of the Yukawa fluid for several strongly coupled state points and compared these with the data from Monte Carlo simulations~\cite{Meijer} and with our previous approximation.~\cite{PractUP} The results are summarized in Table~\ref{Tab2}. Both theoretical approaches demonstrate excellent agreement with MC simulations, especially in the vicinity of the fluid-solid phase transition. Some advantage of the present construction is that the explicit expression for the free energy is available. The validity of the present approach is merely limited by the validity of RT scaling, i.e. to the regime of strong coupling and weak screening.

\begin{table}
\caption{\label{Tab2} The compressibility factor (reduced pressure)  of the single component Yukawa fluid (note, however, that the first row may correspond to supercooled liquid) at strong coupling. The first three columns are similar to those in Tab.~\ref{Tab1}. The remaining columns contain the values of $Z$ obtained using MC simulations by Meijer and Frenkel~\cite{Meijer} and tabulated in Ref.~\onlinecite{Tejero} ($Z_{\rm MC}$), present theoretical approach ($Z_{\rm present}$), and our previous approximation ($Z_{\rm prev}$) from Ref.~\onlinecite{PractUP}.}
\begin{ruledtabular}
\begin{tabular}{lllrrr}
$\kappa$ & $\Gamma$ & $\Gamma/\Gamma_{\rm m}$ &  $Z_{\rm MC}$ & $Z_{\rm present}$ & $Z_{\rm prev}$ \\ \hline
1.800 & 396.9 & 1.03 & 102.492 & 102.498 & 102.526 \\
1.860 & 383.9 & 0.95 & 89.606 & 89.542 & 89.567    \\
1.923 & 371.4 & 0.87 & 78.148 & 78.123 & 78.145   \\
1.984 & 360.0 & 0.80 & 68.640 & 68.618 & 68.637  \\
2.049 & 348.6 & 0.73 & 59.889 & 59.880 & 59.895  \\
2.117 & 337.5 & 0.66 & 52.133 & 52.138 & 52.150  \\
2.182 & 327.3 & 0.60 & 45.711 & 45.699 & 45.707  \\
2.238 & 319.2 & 0.56 & 41.041 & 40.998 & 41.002  \\
2.306 & 309.7 & 0.51 & 35.954 & 35.904 & 35.903  \\
2.348 & 304.2 & 0.48 & 33.204 & 33.189 & 33.184  \\
2.398 & 297.9 & 0.45 & 30.294 & 30.258 & 30.249  \\
2.532 & 282.1 & 0.37 & 23.780 & 23.760 & 23.741  \\
2.631 & 271.5 & 0.32 & 20.016 & 20.017 & 19.989  \\
2.778 & 257.1 & 0.26 & 15.705 & 15.724 & 15.682  \\
\end{tabular}
\end{ruledtabular}
\end{table}

\section{Discussion and Conclusion}

A simple practical approach to estimate thermodynamic properties of strongly coupled Yukawa systems, both in the fluid and solid phases is presented. The excess free energy is obtained by a standard integration of the expressions for the excess internal energy using the RT scaling in the dense fluid phase and harmonic lattice approximation in the solid phase. The integration starts from the freezing/melting point (where the fluid and solid free energies are equal by the conventional definition of this phase transition in Yukawa systems) and requires the knowledge about the location of this point and excess free energy at this point. This information is available from earlier numerical simulations. The accuracy of the approach is verified by calculating the compressibility factors in a wide parameter regime and comparing the results with those from direct MD and MC computer simulations. The relative deviations are documented to be less than $1\%$ for solids in the near-melting and superheated regimes and more than one order of magnitude smaller in other regimes investigated. Given the simplicity and accuracy of the approach, it represents very convenient practical tools to estimate thermodynamic properties of Yukawa systems in a very broad range of parameters.

\begin{figure}
\includegraphics[width=7.5cm]{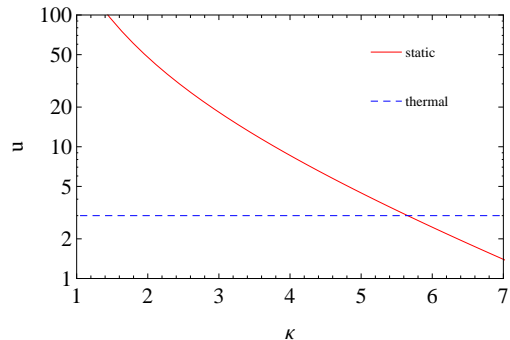}
\caption{(Color online) Approximate static (red solid curve) and thermal (blue dashed curve) components of the reduced excess energy at freezing of Yukawa fluids. The static contribution dominates for $\kappa\lesssim 5$.}
\label{fig3}
\end{figure}

The important reason behind the success of the present approximation is that the static component of the internal energy dominates over the thermal component for soft repulsive interactions. This ensures that some deviations from the quasi-universal behavior of $u_{\rm th}$ for Yukawa fluids lead only to very minor corrections of the total internal energy. This also guarantees that anharmonic corrections to $u_{\rm th}$ in the solid phase do not play very important role. To get an idea how soft the repulsion should be, we plot in Fig.~\ref{fig3} the static component of the internal energy at freezing, $M_{\rm f}\Gamma_{\rm m}$, along with the rough estimate of the thermal energy at freezing,  $u_{\rm th}\simeq 3.0$, as functions of the screening parameter $\kappa$. The two curves intersect around $\kappa\simeq 5.5$. Thus, the absolute upper boundary of the approach applicability can be estimated as $\kappa_{\rm max}\simeq 5$. For larger $\kappa$ the thermal energy plays significant or dominant role, the RT scaling in its simple form used here is no more applicable (see Fig.~\ref{fig2}), anharmonic corrections can play certain role.

Taking into account certain universal scalings in the behavior of simple condensed matter systems~\cite{Saija,Dyre,KhrapakPRL,Univers} we can suggest a more general criteria of the applicability of the approach described in this paper. We conjecture that it should (possibly with some minor modifications) be applicable to soft repulsive systems with the generalized softness parameter $s$ above certain threshold. The generalized softness parameter introduced in Ref.~\onlinecite{Univers} reads
\begin{equation}\label{softness}
s=\left[-1-V''(\Delta)\Delta/V'(\Delta)\right]^{-1},
\end{equation}
where $\Delta$ characterizes the mean interparticle distance and $V(r)$ is the pairwise interaction potential. For example, for the inverse-power law (IPL) interactions $V(r)\propto r^{-n}$ we get the conventional definition $s=1/n$. Using the Yukawa potential and adopting $\kappa_{\rm max}\simeq 5$ we obtain the threshold $s_*\simeq 0.19$. This implies for instance, that thermodynamics of the  IPL systems with $n\lesssim 5$ can be treated in a manner similar to that applied here to Yukawa systems. Indeed, the RT scaling [Eq.~(\ref{fit}) with constant $\delta$] works well for IPL with $n=4$, is slightly less appropriate for $n=6$, and significant deviations are observed for $n=9$ and $n=12$ (see Fig. 1 of Ref.~\onlinecite{Rosenfeld2000}). We expect that the present approach can also be helpful in connection with the thermodynamics of systems exhibiting water-like anomalies (Gaussian core, exp-6, Hertz and related model potentials) in the regime of sufficiently soft interaction (high density), but this requires proper verification. Another interesting question is to which extent the shortest-graph method is applicable for these model systems taking into account their rich solid polymorphism. This is also left for future studies.

The last remark concerns the relevance of the single component Yukawa model to real systems, like complex (dusty) plasmas or colloidal suspensions. First of all, the effect of neutralizing medium has to be properly taken into account. This should not constitute a major problem as we briefly discussed in Sec.~\ref{TF}. Second, the idealized model considered here does not account for some important properties of real systems. For example, the openness of complex plasma systems is known to be responsible for some deviations from the model Yukawa interaction (as was pointed out in the Introduction) and variability of the particle charge. These issues have been repeatedly discussed in recent publications and we refer the reader to Refs.~\onlinecite{DHH,ISM,DAV}. It is not clear at the moment how large the related modifications to thermodynamic properties can be and whether they can be evaluated using conventional methods. Nevertheless, the present approach can be an important element in constructing more realistic and accurate models for the thermodynamics of real systems.

\begin{acknowledgments}
This study has been supported by the Russian Science Foundation, Project No. 14-43-00053.
The numerical simulations have been supported by the Russian Ministry of Education and Science, Project No. 3.1526.2014/K.
At a late stage of this work SAK has been supported by the A*MIDEX grant (Nr.~ANR-11-IDEX-0001-02) funded by the French Government ``Investissements
d'Avenir" program.
\end{acknowledgments}


\begin{thebibliography}{99}

\bibitem{Hopkins} P. Hopkins, A. J. Archer, and R. Evans, J. Chem. Phys. {\bf 124}, 054503 (2006).

\bibitem{Lowen} H. L\"{o}wen, Phys. Rep. {\bf 237}, 249 (1994).
\bibitem{Morrison} I. D. Morrison and S. Ross, {\it Colloidal Dispersions: Suspensions, Emulsions, and Foams} (Wiley, 2002).
\bibitem{IvlevBook} A. Ivlev, H. L\"{o}wen, G. Morfill, and C. P. Royall, {\it Complex Plasmas and Colloidal Dispersions: Particle-resolved Studies of Classical Liquids and Solids} (World Scientific, Singapore, 2012).

\bibitem{FortovRev} V. E. Fortov, A. G. Khrapak, S. A. Khrapak, V. I. Molotkov, and O. F. Petrov,
Phys. Usp. {\bf 47}, 447 (2004); V. E. Fortov, A. V. Ivlev, S. A. Khrapak, A. G. Khrapak, and G. E. Morfill, Phys. Rep. {\bf 421}, 1 (2005).
\bibitem{ShuklaRMP} P. K. Shukla and B. Eliasson, Rev. Mod. Phys. {\bf 81}, 25 (2009).
\bibitem{FortovBook} {\it Complex and dusty plasmas: From Laboratory to Space}, edited by V. E. Fortov and G. E. Morfill (CRC Press, Boca Raton, 2010).
\bibitem{ChaudhuriSM} M. Chaudhuri, A. V. Ivlev, S. A. Khrapak, H. M. Thomas, G. E. Morfill, Soft Matter {\bf 7}, 1287 (2011).

\bibitem{KhrapakPRL2008} S. A. Khrapak, B. A. Klumov, G. E. Morfill, Phys. Rev. Lett. {\bf 100}, 225003 (2008).
\bibitem{KhrapakCPP} S. Khrapak and G. Morfill, Contrib. Plasma Phys. {\bf 49}, 148 (2009).
\bibitem{Semenov} I. L. Semenov, A. G. Zagorodny, and I. V. Krivtsun, Phys. Plasmas {\bf 18}, 103707 (2011).

\bibitem{KhrapakPoP2010} S. A. Khrapak, A. V. Ivlev, and G. E. Morfill, Phys. Plasmas {\bf 17}, 042107 (2010).
\bibitem{Wysocki} A. Wysocki, C. R\"{a}th, A. V. Ivlev, K. R. S\"{u}tterlin, H. M. Thomas, S. Khrapak, S. Zhdanov, V. E. Fortov,
A. M. Lipaev, V. I. Molotkov, O. F. Petrov, H. L\"{o}wen, and G. E. Morfill, Phys. Rev. Lett. {\bf 105}, 045001 (2010).

\bibitem{Robbins} M. O. Robbins, K. Kremer, and G. S. Grest, J. Chem. Phys. {\bf 88}, 3286 (1988).
\bibitem{Meijer} E. J. Meijer and D. Frenkel, J. Chem. Phys. {\bf 94}, 2269 (1991).
\bibitem{Hamaguchi96} S. Hamaguchi, R. T. Farouki, and D. H. E. Dubin, J. Chem. Phys. {\bf 105}, 7641 (1996).
\bibitem{Hamaguchi97} S. Hamaguchi, R. T. Farouki, and D. H. E. Dubin, Phys. Rev. E {\bf 56}, 4671 (1997).
\bibitem{CG} J. M. Caillol and D. Gilles, J. Stat. Phys. {\bf 100}, 933 (2000).

\bibitem{Tejero} C. F. Tejero, J. F. Lutsko, J. L. Colot, and M. Baus, Phys. Rev. A {\bf 46}, 3373 (1992).
\bibitem{Kalman2000} G. J. Kalman, M. Rosenberg, and H. DeWitt, J. Phys. IV France {\bf 10}, 403 (2000).
\bibitem{Faussurier} G. Faussurier, Phys. Rev. E {\bf 69}, 066402 (2004).
\bibitem{Gapinski} J. Gapinski, G. N\"{a}gele, and A. Patkowski, J. Chem. Phys. {\bf 136}, 024507 (2012); {\bf 141}, 124505 (2014).
\bibitem{Hlushak} S. Hlushak and A. Trokhymchuk, Condens. Matter Phys. {\bf 15}, 23003 (2012); S. Hlushak, J. Chem. Phys. {\bf 141}, 204108 (2014).
\bibitem{SMSA} P. Tolias, S. Ratynskaia, and U. de Angelis, Phys. Rev. E {\bf 90}, 053101 (2014).

\bibitem{Com1} We consider only three-dimensional (3D) case in this study. There exists also extensive literature on the thermodynamics of 2D Yukawa systems, which is not included in this reference list.

\bibitem{TotsujiJPA} H. Totsuji, J. Phys. A: Math. Gen. {\bf 39}, 4565 (2006).
\bibitem{TotsujiPoP} H. Totsuji, Phys. Plasmas {\bf 15}, 072111 (2008).
\bibitem{Vaulina2010} O. S. Vaulina, X. G. Koss, Yu. V. Khrustalev, O. F. Petrov, and V. E. Fortov, Phys. Rev. E {\bf 82}, 056411 (2010).

\bibitem{DHH} S. A. Khrapak, A. G. Khrapak, A. V. Ivlev, and G. E. Morfill, Phys. Rev. E {\bf 89}, 023102 (2014).
\bibitem{ISM} S. A. Khrapak, A. G. Khrapak, A. V. Ivlev, and H. M. Thomas, Phys. Plasmas {\bf 21}, 123705 (2014).

\bibitem{Yurchenko} S. O. Yurchenko, J. Chem. Phys. {\bf 140}, 134502 (2014).
\bibitem{YurchenkoPRB} S. O. Yurchenko, N. P. Kryuchkov, and A. V. Ivlev, J. Chem. Phys. (2015, submitted).

\bibitem{Landau} L. D. Landau and E. M. Lifshitz, {\it Statistical Physics} (Elsevier, 2005).

\bibitem{Ham94} S. Hamaguchi and R. T. Farouki, J. Chem. Phys. {\bf 101}, 9876 (1994).

\bibitem{Ross} M. Ross, Phys. Rev. {\bf 184}, 233 (1969).
\bibitem{Hoover} W. G. Hoover, S. G. Gray, and K. W. Johnson, J. Chem. Phys. {\bf 55}, 1128 (1971).
\bibitem{Agrawal} R. Agrawal and D. A. Kofke, Mol. Phys. {\bf 85}, 23 (1995).


\bibitem{VaulinaJETP} O. S. Vaulina and S. A. Khrapak, JETP {\bf 90}, 287 (2000).
\bibitem{VaulinaPRE} O. Vaulina, S. Khrapak, and G. Morfill, Phys. Rev. E {\bf 66}, 016404 (2002).

\bibitem{PractUP} S. A. Khrapak and H. Thomas, Phys. Rev. E {\bf 91}, 023108 (2015).

\bibitem{HansenBook} J. P. Hansen and I. R. McDonald, {\it Theory of Simple Liquids} (Academic Press, 2006).

\bibitem{Rosenfeld1998} Y. Rosenfeld and P. Tarazona, Mol. Phys. {\bf 95}, 141 (1998).
\bibitem{Rosenfeld2000} Y. Rosenfeld, Phys. Rev. E {\bf 62}, 7524 (2000).

\bibitem{Ingebrigtsen} T. S. Ingebrigtsen, A. A. Veldhorst, T. B. Schr{\o}der, and J. C. Dyre, J. Chem. Phys. {\bf 139}, 171101 (2013).

\bibitem{Farouki} R. T. Farouki and S. Hamaguchi, J. Chem. Phys. {\bf 101}, 9885 (1994).

\bibitem{Hoy} R. S. Hoy and M. O. Robbins, Phys. Rev. E {\bf 69}, 056103 (2004).
\bibitem{KhrapakEPL2012} S. A. Khrapak and G. E. Morfill, EPL {\bf 100}, 66004 (2012).

\bibitem{Saija} F. Saija, S. Prestipino, and P. V. Giaquinta, J. Chem. Phys. {\bf 115}, 7586 (2001).
\bibitem{Dyre} J. C. Dyre, J. Phys. Chem. B {\bf 118}, 10007 (2014).
\bibitem{KhrapakPRL} S. A. Khrapak and G. E. Morfill, Phys. Rev. Lett. {\bf 103}, 255003 (2009).
\bibitem{Univers} S. A. Khrapak, M. Chaudhuri, and G. E. Morfill, J. Chem. Phys. {\bf 134}, 241101 (2011).


\bibitem{DAV} S. A. Khrapak and H. M. Thomas, Phys. Rev. E {\bf 91}, 033110 (2015).

\end{thebibliography}
\end{document}